\documentclass[useAMS,usenatbib]{mn2e}

\usepackage{graphicx}
\usepackage{comment}
\usepackage{color}
\usepackage{subfig} 
\title[Radiative Shocks around Super-Eddington Accreting Black Holes]
 {Radiative Shocks around Super-Eddington Accreting  Black Holes}
\author[T. Okuda, C. B. Singh ]
 {Toru Okuda$^{1}$ \thanks{E-mail:bbnbh669@ybb.ne.jp},
 Chandra B. Singh$^{2}$  \\ 
 $^{1}$ Hakodate Campus, Hokkaido University of Education, Hachiman-Cho 1-2, 
  Hakodate 040-8567, Japan \\
 $^{2}$  South-Western Institute for Astronomy Research, Yunnan University, University Town,
  Chenggong, \\
\:\: Kunming 650500, People's Republic of China \\
  }

\usepackage{amsmath}

\begin{document}

\date{Accepted }

\pagerange{\pageref{0}--\pageref{0}} \pubyear{2012}

\maketitle

\label{firstpage}

\begin{abstract}
 We examine radiative standing shocks in advective accretion flows 
 around  stellar-mass black holes by 2D radiation hydrodynamic simulations, 
 focusing on the super-Eddington
 accreting flow.  Under a set of input flow parameters responsible for the standing shock, 
 the shock location on the equator decreases toward the event horizon with an increasing 
 accretion rate. The optically thin and hot gas in the narrow funnel region along
 the rotational axis changes gradually into a dense and optically thick state with 
 the increasingly dense gas transported from the base of the radiative shock 
 near the equator. As a result, the  luminosity becomes as high as $\sim 10^{40}$ 
 erg s$^{-1}$, and the radiation shows a strongly anisotropic distribution around 
 the rotational axis and then very low edge-on luminosity as $\sim 10^{36}$ erg s$^{-1}$. 
 The mass outflow rate from the outer boundary is high as 
 $\sim 10^{-5}$ and $10^{-4} M_{\odot}$ yr$^{-1}$ but  most of the outflow  is 
 originated through the radial outer  boundary and may be observed over a wide wind region.
The models show approximately black body spectra with a temperature of
$5 \times 10^6$ -- $3 \times 10^7$ K at the vertical outer boundary surface.
  The radiative shock models with the super-Eddington luminosities show a possible model 
 for the superaccretor SS 433 and Ultraluminous X-ray sources with stellar-mass black holes.
\end{abstract}

\begin{keywords}
accretion, accretion discs -- black hole physics  -- 
radiation mechanism: thermal -- shock waves.

\end{keywords}

\section{Introduction}
Advective accretion flow around a black hole is likely to have two saddle-type sonic points. After the accretion flow with the appropriate injection parameters passes through the outer sonic points, the supersonic flow can be virtually stopped by the centrifugal force, forming a standing shock close to the black hole and again falling into the black hole supersonically. Apart from general researches on the shock wave,  the study of the standing shock in the accretion disc flow under an astrophysical context was pioneered by
 \citet {b13}. He examined the transonic disc accretion onto a black hole with a fully relativistic treatment, assuming constant disc height and found multiple critical points as well as a transition through standing shock. \citet{b4} considered the hydrostatic equilibrium in the vertical direction and presented analytically the transonic solutions which include the standing shock in the rotating adiabatic flow under the pseudo-Newtonian potential of the black hole.
Further studies of the standing shocks in the low angular momentum flows have been developed  
 for parameter space research  of the specific internal energy  $\mathcal{E}$
 and the specific angular momentum $\lambda$ responsible for the standing shock \citep{b10, b8, b27}, for 2D numerical simulations of the shocks by smoothed particle hydrodynamics (SPH) or Eulerian 2D hydrodynamics \citep{b24, b25, b26, b22, b18}, for astrophysical applications such as hard X-ray emission from the post-shock region, mass outflow rate and  quasi-periodic oscillations (QPOs) phenomena \citep{b6, b7, b38, b30} and for effects of cooling, viscosity, and magnetic field on the standing shock \citep{b9, b21, b36, b1, b32}.
However, there are only several 1D theoretical \citep{b14, b15, b16, b17} and 2D numerical studies   \citep{b34, b33} of radiative standing shock, where the radiation influences on the shock structure of the pre-shock and the post-shock regions. 

 In this paper, we examine further radiative shocks in the accretion flow by 2D radiation hydrodynamic
 simulations, especially focusing on the super-Eddington accreting flow, compare the results with
 theoretical structure of the radiative shock by \citet{b41} and  \citet{b16}, and discuss the characteristic features of the radiative shock models in relevance
 to Ultraluminous X-ray sources (ULXs). In the next section, we present the model equations
 of our simulation setup, then numerical schemes, initial and boundary conditions are
 mentioned in section 3. In section 4, the results along with astrophysical significance
 are discussed then we summarize and discuss in the final section.

\section{Model equations}

 We examine these shock problems in two-dimensional inviscid flow.
 A set of relevant equations of six partial differential equations of density, momentum, and thermal and radiation energy are solved.
 These equations include the heating and cooling of gas and radiation transport \citep{b20}. 
Using cylindrical coordinate ($r$, $z$, $\phi$), the basic equations are written in the following conservation form:

\begin{equation}
 {\partial \rho\over\partial t} + \nabla\cdot\left(\rho\textbf{v}\right) = 0, 
\end{equation}

\begin{equation}
\begin{split}
%\begin{equation}
 {\partial (\rho{v})\over\partial t} + \nabla\cdot\left(\rho v\textbf{v}\right)
  &=  \rho\left[{{v_{\phi}}^2\over r} - {GM_{*}\over (\sqrt{r^2 + z^2}-r_{\rm g})^2} {r\over \sqrt{r^2+z^2}} \right] \\
  &-{\partial p\over\partial r} + f_{r}, 
\end{split}
\end{equation}
%\end{equation}

\begin{equation}
 {\partial (\rho w)\over\partial t} + \nabla\cdot\left(\rho w\textbf{v}\right)
  = - {\rho GM_{*}\over (\sqrt{r^2 + z^2}-r_{\rm g})^2} {z\over \sqrt{r^2+z^2}} 
  -{\partial p\over\partial z} + f_{z},
\end{equation}
\begin{equation}
 {\partial (\rho r v_{\phi})\over\partial t} + \nabla\cdot\left(\rho r v_{\phi}\textbf{v}\right)
  = 0,
\end{equation}
\begin{equation}
 \frac{\partial{\rho\epsilon}}{\partial t} + \nabla\cdot \left(\rho\epsilon\textbf{v}\right)
   =  -p \nabla \textbf{v} - \Lambda,
\end{equation}
\begin{equation}
 \frac{\partial E_{0}}{\partial t} +\nabla\cdot {\textbf{F}}_{0} + \nabla\cdot (\textbf{v}E_{0}+\textbf{v}\cdot P_{0})
  = \Lambda - \rho{(\kappa+\sigma)\over c} \textbf{v}\cdot {\textbf{F}}_{0},
\end{equation}
where $\rho$ is the mass density, $\textbf{v}= (v, w, v_{\phi})$ is the three fluid velocity components, 
$G$ is the gravitational constant, $M_{*}$ is the black hole mass, $p$ is the gas pressure, $\epsilon$ is the specific internal energy, $E_{0}$ is the radiation energy density per unit volume, $ {\textbf{P}}_{0}$ is the radiative stress
tensor and $c$ is the speed of light. Here, subscript ``0" denotes the value in the comoving frame \citep{b19} and a pseudo-Newtonian potential \citep{b35} is adopted, where $r_{\rm g}$ is the Schwarzschild radius, given by $2GM_{*}/c^2$. The force density,  ${\textbf{f}}_{\rm R} = (f_{r}, f_{z})$,
 exerted by the radiation field is given by

\begin{equation}
   {\textbf{f}}_{\rm R} = \rho {(\kappa+\sigma) \over c} {\textbf{F}}_{0},
\end{equation}
where $\kappa$ and $\sigma$ denote the absorption and scattering coefficient and ${\textbf{F}}_{0}$ is the radiative
flux in the comoving frame.

 The quantity $\Lambda$ denotes the cooling and heating rates of the gas,
\begin{equation}
 \Lambda = \rho c\kappa(S_{*} - E_{0}),
\end{equation}
where $S_{*}$ is the source function. For this source function, we assume local thermal equilibrium, $S_{*}=aT^{4}$,
where $T$ is the gas temperature and $a$ is the radiation constant. For the equation of state, the gas pressure is given by the ideal gas law, $p = R_{\rm G}\rho T/\mu$, where $\mu$ is the mean molecular weight and $R_{\rm G}$ is the gas constant. To close the system of equations, we use the flux-limited diffusion approximation \citep{b23} for the radiative flux:

\begin{equation}
  {\textbf{F}}_{0} = - {{\lambda_{\rm f} c}\over{\rho(\kappa + \sigma)}} \nabla {E}_{0}.
\end{equation}

and

\begin{equation}
  {\textbf{P}}_{0} = E_{0}\cdot T_{\rm Edd},
\end{equation}
where $\lambda_{\rm f}$ and $T_{\rm Edd}$ are the {\it flux-limiter} and the {\it Eddington Tensor}, respectively, for which we use the approximate formulas given in \citet{b20}.

\section{Numerical methods}
\subsection{Model parameters and numerical scheme}
For the central black hole, we assume a Schwarzschild black hole with mass $M_{*}$ = 10$M_{\odot}$.
First, we set model parameters of a specific internal energy $\mathcal{E}$
 and a specific angular momentum $\lambda$
 which are responsible for the standing shock.
 Then, given an outer radial boundary $R_{\rm out}$ and an input accretion rate 
 $\dot M_{\rm input}$, we obtain flow variables of the density $\rho_{\rm out}$, the radial velocity $v_{\rm out}$
 and the sound velocity $a_{\rm out}$ at the outer radial boundary from 1.5D transonic solutions \citep{b4}, 
 under the vertical hydrostatic equilibrium assumption of the flow.  For the given sound velocity $a_{\rm out}$, the gas temperature
 $T_{\rm out}$ is obtained from $a_{\rm out}^{2} ={\gamma R_{\rm G}T_{\rm out}} /\mu$ and $\gamma aT_{\rm out}^{4}/ {3{\rho}_{\rm out}}$
 for gas-pressure dominant and radiation-pressure dominant cases, respectively, where $\gamma$ is the specific heat ratio. 
Accordingly, if the input gas is gas-pressure dominant, 
 the temperature $T_{\rm out}$ is determined independent of input density $\rho_{\rm out}$. On the other hand,  if the radiation pressure is dominant, the input gas temperature $T_{\rm out}$ and the radiation energy density $(E_{0})_{\rm out} (= a T_{\rm out}^{4})$ 
increases with increasing input density,  assuming the local thermal equilibrium at the outer boundary.
In 1.5D transonic equations, we have the specific energy  $\mathcal{E}$
, that is, Bernoulli constant, given by

\begin{equation}\label{eq:tot_E}
   \mathcal{E}
 = \frac {{v}_{\rm out}^{2}} {2} + \frac {a_{\rm out}^{2}} {\gamma - 1} + \frac {\lambda^{2}}{2r_{\rm out}^{2}} -\frac {1} {2(r_{\rm out}-1)},
\end{equation}
where  hereafter the velocities $v_{\rm out}$ and $a_{\rm out}$, $r$ and $\lambda$ are given by units of $c$, $r_{\rm g}$ and $r_{\rm g} c$, respectively.  The specific heat ratio 
$\gamma$ of 1.6 is taken throughout this paper.

We examine the standing shock in the flow with a fixed set of $\mathcal{E}$
 and $\lambda$. As is mentioned above, when the gas pressure is dominant,  $T_{\rm out}$ is specified by the fixed  $\mathcal{E}$ and $\lambda$, but
 when the radiation-pressure is dominant, we need to specify furthermore the input density $\rho_{\rm out}$ to get the outer
 boundary temperature $T_{\rm out}$.  In the flux-limited diffusion approximation, the radiative flux $F_0$ is given by 
the flux limiter $\lambda_{\rm f}$, the flux limiter $\lambda_{\rm f}$ depends on $E_0$,
 $\rho$ and  $\kappa$, and achieves an extreme limit of 1/3 if the local density is sufficiently large, that is, the gas is fully  optically thick.
 Then, the flux-limited diffusion approximation becomes the Eddington diffusion approximation.
Thus the input density $\rho_{\rm out}$ is a parameter that evaluates the degree of optical thickness of the flow.

We take the outer boundary $R_{\rm out}$ = 100.
 Referring to the hydrodynamical results in our previous paper \citep{b32},
we adopt  $\lambda$= 1.5, $v_{\rm out}$= -0.072241 and $a_{\rm out}$= 0.037566 
 at the outer radial  boundary.
Then, giving the input density $\rho_{\rm out}$, we get other variables at the radial outer boundary.
We examine six models of  adiabatic case with $\rho_{\rm out} =2 $ (AD), optically thin case with $\rho_{\rm out}$ =2 (Thin1) and four optically thick cases with $\rho_{\rm out}$ = 20,
 200, 2$\times 10^3$ and 2$\times 10^4$ (Thick2, Thick3, Thick4, and Thick5), respectively.
Hereafter the density is shown in unit of  ${\rho}_0 =10^{-8}$ g cm$^{-3}$.
In model Thin1, the input gas is given to be optically thin but in models 
Thick2 to Thick5 the input accreting gas
is optically thick. We should notice that  model Thin1 has the same
 parameters as the adiabatic case AD but in the latter case, the radiation transport is not treated.
The flow parameters and other input variables at the outer radial boundary are listed in Table 1,
 where $\mathcal{E}$= 2.328 $\times 10^{-5}$ and $\lambda$= 1.5 are used.

\begin{table*}
\centering
\caption{Flow arameters of the radial velocity $v_{\rm out}$, the sound velocity $a_{\rm out}$, the input density $\rho_{\rm out}$, the temperature $T_{\rm out}$, the radiation energy density $(E_0)_{\rm out}$, the input accretion rate $\dot M_{\rm input}$ at the outer radial boundary $R_{\rm out}$= 100, where
  $\mathcal{E}$= 2.328$\times 10^{-5}$ and $\lambda$= 1.5 are used and
  $\rho_{0}$ is $10^{-8}$ g  cm$^{-3}$.}
 . . 
\vspace{3mm}
\begin{tabular}{@{}ccccccc} \hline
&&&&&& \\

${\rm model}$ & $v_{\rm out}$ & $a_{\rm out}$ & $\rho_{\rm out}$ & $T_{\rm out}$ 
& $(E_0)_{\rm out}$ & ${\dot M}_{\rm input}$  \\
  & ($c$) & ($c$) & (${\rho}_0$) & ($ \rm K$) &  (${\rho}_0 c^2$) 
 & (${\dot M}_{\rm E}$)  \\
 \hline

 ${\rm AD}$        & -7.224E-2 & 3.757E-2 & 2    & 4.767E9 & ---  & 12 \\
 ${\rm Thin1}$     &   ''          &      ''        & 2    & 4.767E9 & 5.173E-5  & 12 \\ 
 ${\rm Thick2}$   &    ''         &      ''        & 20   & 2.816E6 & 5.292E-2 & 120 \\
 ${\rm Thick3}$   &    ''         &      ''        & 200  & 5.007E6 & 5.292E-1& 1.2E3\\
 ${\rm Thick4}$   &    ''         &      ''        & 2E3  & 8.905E6 & 5.292      &1.2E4 \\
 ${\rm Thick5}$   &    ''         &     "         & 2E4  & 1.583E7 & 5.292E1  & 1.2E5 \\
\hline
\end{tabular}
\end{table*}
 
Here we introduce the Eddington critical accretion rate ${\dot M}_{\rm E} (=L_{\rm E}/c^2)$, where $L_{\rm E}$ is the
Eddington luminosity given by,

\begin{equation}
  L_{\rm E} = \frac{4\pi GM_*c}{\kappa_{\rm e}},
\end{equation}
where ${\kappa}_{\rm e}$ is the electron scattering opacity. $L_{\rm E}$ and ${\dot M}_{\rm E}$ are 1.5 $\times 10^{39}$ erg s$^{-1}$ and 1.7$\times 10^{18}$, respectively, for the stellar-mass black hole with $M_* = 10 M_{\rm \odot}$.

\begin{table*}
\centering
\caption{The shock location $R_{\rm s}$ on the equator, the total luminosity $L$, the luminosity $L_{\rm f}$ from the funnel region, the luminosity $L_{\rm rout}$ from the radial outer boundary, 
 the total mass outflow rate ${\dot M}_{\rm out}$, the mass inflow rate ${\dot M}_{\rm edge}$ 
at the inner edge of the flow, and the mass outflow rate ${\dot M}_{\rm rout}$ through the outer radial boundary
 obtained in the simulations.}
\vspace{3mm}
\begin{tabular}{@{}cccccccc} \hline
&&&&&&& \\

${\rm model}$ & $R_{\rm s}$ & $L$& $L_{\rm f}/L$ &  $L_{\rm rout}/L$ & ${\dot M}_{\rm out}$ 
 & ${\dot M}_{\rm edge}$ &  ${\dot M}_{\rm rout}/\dot M_{\rm out}$   \\
  & ($r_{\rm g}$) & ($L_{\rm E}$) &  -- & -- & (${\dot M}_{\rm input}$) & 
  (${\dot M}_{\rm input}$) & --   \\
 \hline
 ${\rm AD}$        &  50  & 0.02  & --    & --  &   0.25   &   0.7      & 0.98 \\
 ${\rm Thin1}$     & 50  &  0.02  & 0.2  & 8E-3   &  0.21   &  0.8       & 0.97  \\ 
 ${\rm Thick2}$   & 15  & 0.76   & 0.1   & 8E-4  & 0.04 &   0.9       & 0.95  \\
 ${\rm Thick3}$   & 12  & 4.1     & 0.2   & 9E-4  & 0.02 & $\sim 1$ & 0.97 \\
 ${\rm Thick4}$   &  9  & 8.1      & 0.79 & 7E-5  & 0.04 & $\sim 1$  & 0.77  \\
 ${\rm Thick5}$   &  8  & 18.3    &  0.82 & 7E-5  & 0.04 & $\sim 1$  & 0.80  \\
\hline
\end{tabular}
\end{table*}

\begin{figure}
 \begin{center}
     \includegraphics[width=86mm,height=60mm,angle=0]{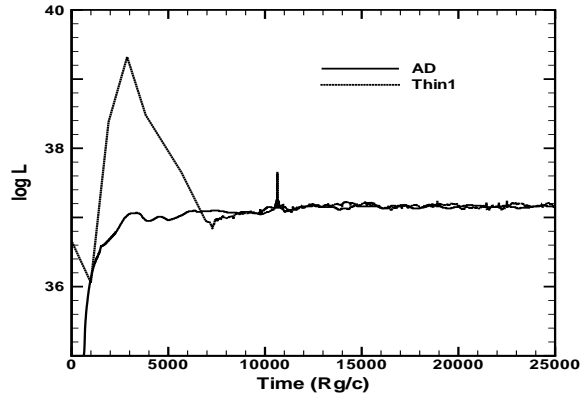}
     \label{fig1}
 \end{center}
%\vspace*{24mm}
 \caption{Time variations of luminosity $L$ (erg s$^{-1}$) for modeld AD (solid line) and Thin1 (dotted line).}
 \end{figure}

\subsection{Initial conditions}
  The 1.5D transonic solutions of the accretion flow give the initial conditions, that is, density $\rho(r)$, radial velocity
 $v(r)$, sound speed $a(r)$, Mach number $M_{\rm a}(r)$, temperature $T(r)$ and disc thickness $h(r)$
 within $ z \leq h(r)$  at a given radius $r$. 
 In the region of  $z  \ge h(r)$, the variables are set appropriately.

\subsection{Boundary Conditions}
  The outer radial boundary 
  at $r=R_{\rm out}$ is divided into two parts. One is the disc outer boundary through which matter is entering 
 from the outer flow.
 At the disc boundary ( $0 \leq z \leq h_{\rm out}$ at $r$ =$R_{\rm out}$), we impose continuous
 inflow of matter with constant variables given by the 1.5D solutions, where $h_{\rm out}$ is the
 disc height at the outer radial boundary.
 The other is the outer boundary region above the disc. Here we impose free-floating conditions and allow
 for the outflow of matter, whereas any inflow is prohibited.
 At the vertical outer boundary $z=Z_{\rm out}$ (=100), we also impose the free floating
 conditions. On the rotating axis, all variables are set to be symmetric relative to the axis.
 The inner boundary at $r = R_{\rm in}$ (=2) is treated as the absorbing boundary since it is below the last stable circular  orbital radius 3$r_{\rm g}$.

\begin{figure}
 \centering
 \subfloat[Profiles of temperature $T$ (K) and Mach number on the equator of the steady flow in model Thin. ]{
 \label{fig2a}
 \includegraphics[width=86mm,height=66mm]{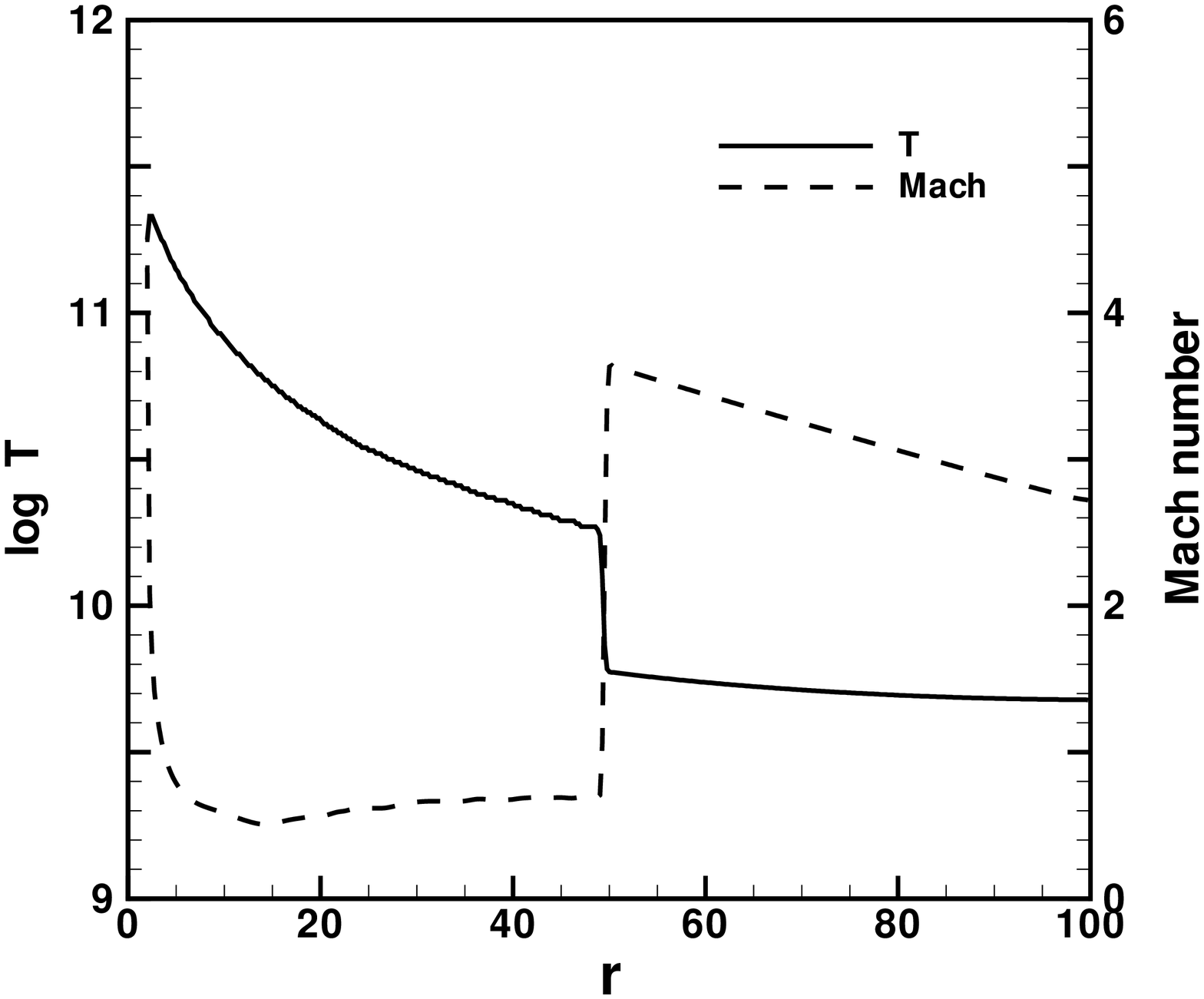} } 
 
\subfloat[2D contours of temperature in model Thin1, where the shock front is shown as the thick solid line elongated upward ($R_{\rm s} \sim 50$).]{
 \label{fig2b}
 \includegraphics[width=86mm,height=66mm]{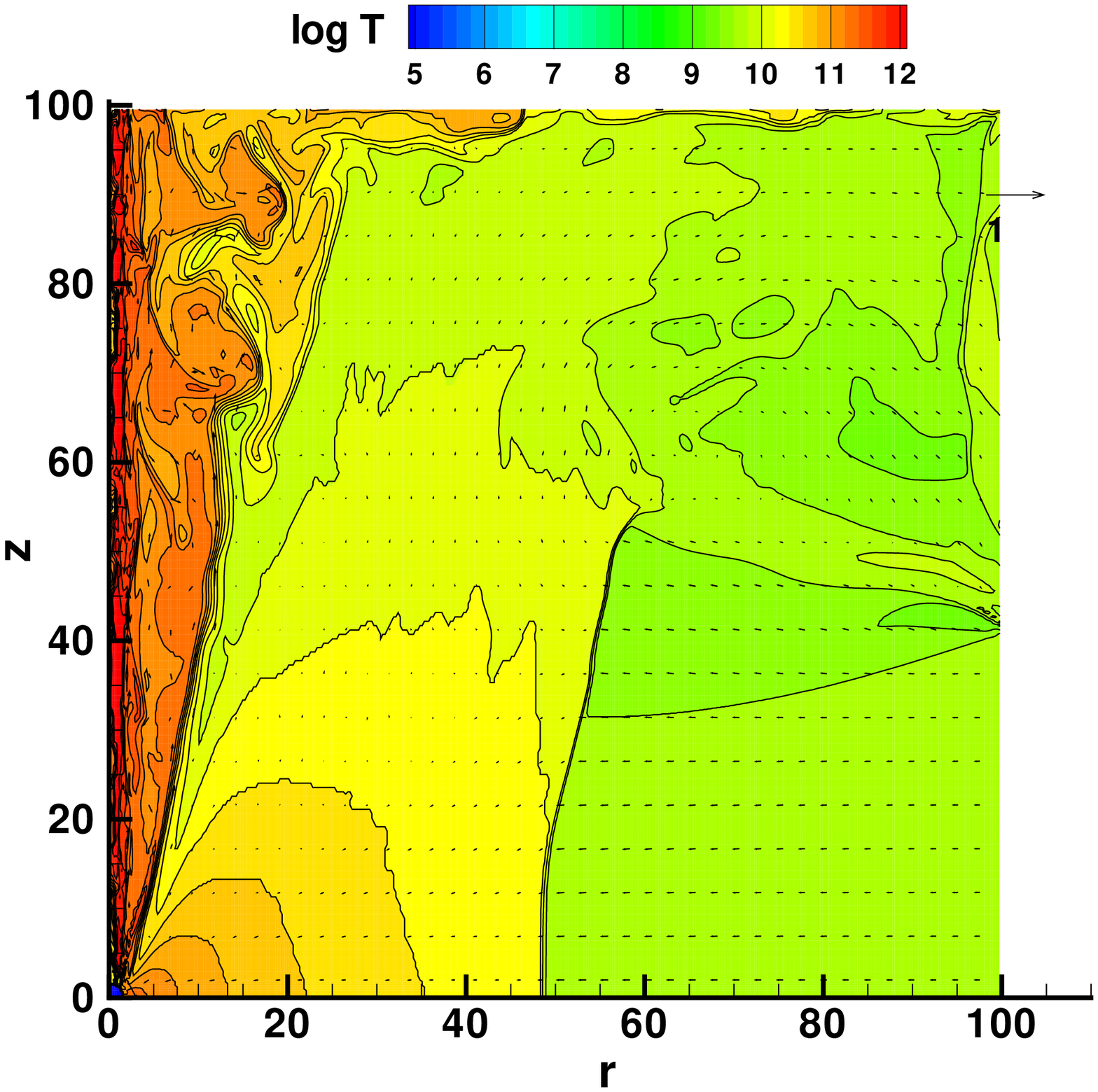} } 
 \caption{}
 \label{fig2}
 \end{figure}

\section{Numerical Results}

The set of partial differential equations (1)-(6) is numerically solved by a finite-difference method under adequate initial and boundary conditions. 
The numerical schemes used are the same as those described previously \citep{b31}. The methods are based on an explicit-implicit finite difference scheme that is superior in numerical stability. The mesh points $N_{\rm r}$ and $N_{\rm z}$ in radial and vertical directions are taken to be 410 and 210, respectively, in all simulations.

\subsection{Luminosity and shock location}
The luminosity in model AD is given by, assuming that the gas is completely
 optically thin 

 \begin{equation}
 L=\int \Lambda {\rm d} V,
\end{equation}
 where only bremsstrahlung emission is considered under one temperature model and $L$ is integrated over all
 computational zones. 
On the other hand, the luminosity  in models Thin1 to Thick5 is given by 

 \begin{equation}
  L = \int \textbf{F}_0 d \textbf{S},
 \end{equation}
where  $d \textbf{S}$ is the area element of the computational
domain and the integral is taken over the surface area of the computational domain except for the region of the input accreting gas.
The mass outflow rate $\dot M_{\rm out}$ is defined by the total rate of outflow
 through the outer boundaries ($z= Z_{\rm out}$) in the z-direction and   ($r= R_{\rm out}$) in the r-direction.

 \begin{equation}
\begin{split}
  \dot M_{\rm out} =& 4\pi\int_{0}^{R_{\rm out}} \rho (r, Z_{\rm out}) w(r, Z_{\rm out}) r dr \\
      & + 4\pi \int_{h_{\rm out}}^{Z_{\rm out}} R_{\rm out}^2 \rho (R_{\rm out}, z) v(R_{\rm out}, z) dz.
 \end{split}
\end{equation}

\begin{figure}
 \centering
 \subfloat[Profiles of  temperature $T$ (K) and  Mach number of the gas
     on the equator in model Thick3. ]{
 \label{fig3a}
 \includegraphics[width=86mm,height=66mm]{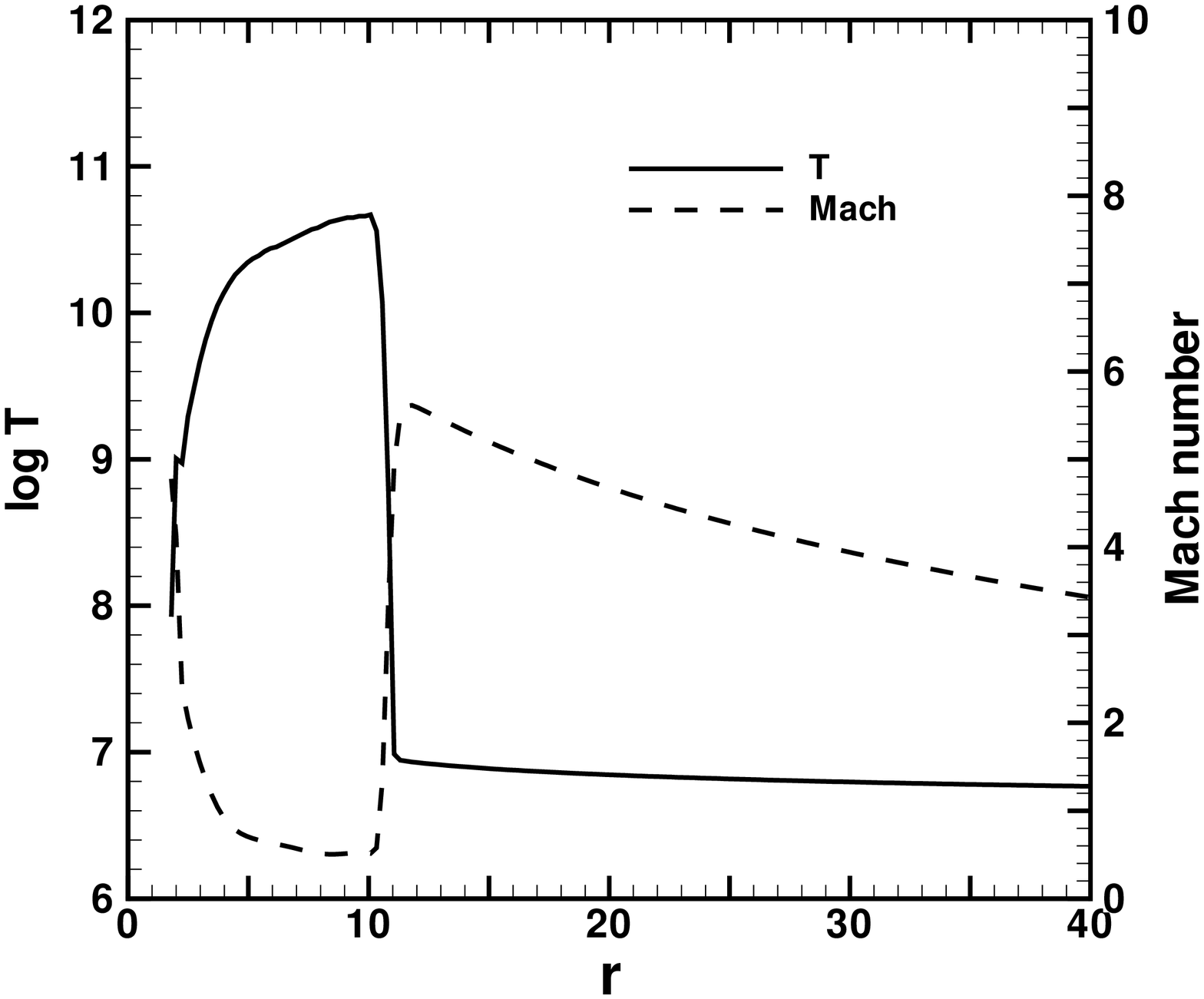} } 
 
\subfloat[2D contours of temperature in model Thick3,
  where the shock front is shown as the thick solid line elongated upward ($R_{\rm s} \sim 10$).]{
 \label{fig3b}
 \includegraphics[width=86mm,height=66mm]{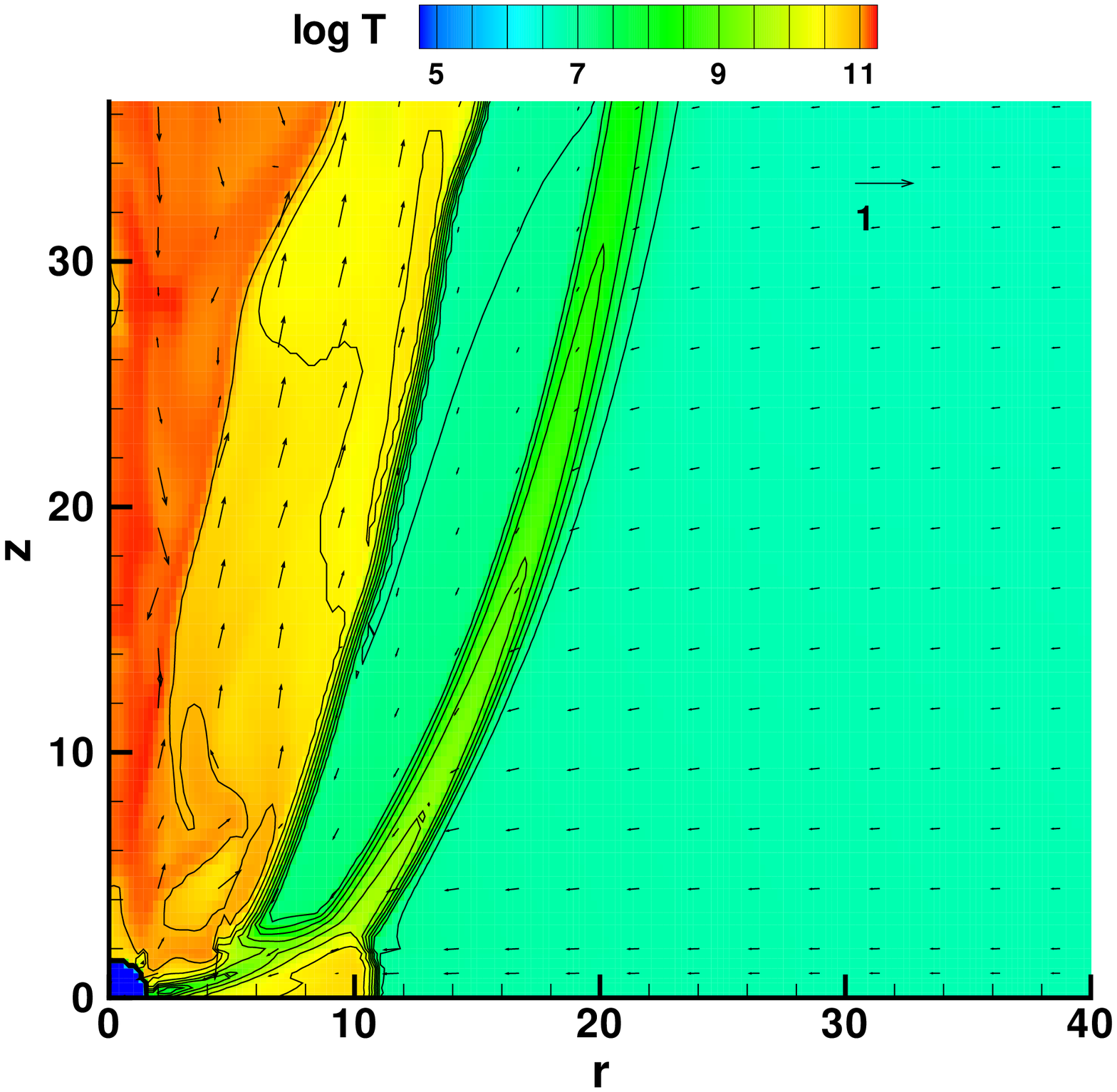} } 
 \caption{}
 \label{fig3}
 \end{figure}

The simulations are done until steady-state flow is obtained.  
Table 2 shows the shock location $R_{\rm s}$ on the equator, the total luminosity $L$,
 the funnel luminosity $L_{\rm f}$ from the funnel region, the edge-on luminosity $L_{\rm rout}$ through the radial outer boundary, the total mass flow rate ${\dot M}_{\rm out}$ from
 the outer boundaries, the mass inflow rate ${\dot M}_{\rm edge}$ at the inner edge of the flow
  and the mass outflow rate $\dot M_{\rm rout}$ through the radial outer boundary. 

Fig.~1 shows the luminosity curves for models AD and Thin1. The luminosity curve in the optically thin model Thin1 agrees well with that in the adiabatic model, where the final luminosity 
 $L \sim 3 \times 10^{37}$ erg s$^{-1}$.
 This result confirms that the flux-limited diffusion approximation used here has a good accuracy even in optically thin gas.
 Fig.~2 (a) shows profiles of the temperature $T$ and the Mach number
 on the equator in model Thin1, which are almost same as those in model AD.
The shock front is found as a sharp discontinuity at $r \sim 50$.
 Fig.~2 (b) shows 2D contours of temperature in model Thin1, where the shock front is shown as a thick solid line elongated upward
 ($R_{\rm s} \sim 50$).  Contours of the flux-limitter $\lambda_{\rm f}$ in the model denote that the gas near the equator becomes optically thick as $\lambda_{\rm f} \sim 0.3$ in the post-shock region but is moderately optically thin as $\lambda_{\rm f} \sim 0.2$ in the upstream and fully optically thin in the funnel region.

On the other hand,  in the radiation-pressure dominant models, the increasing input densities
 $\rho_{\rm out}$ enhance  the optical thickness of the gas. The shock locations in models 
Thick2 to Thick5  decrease as $R_{\rm s} \sim$ 15 to 8, respectively. 
The increasing density $\rho_{\rm out}$ leads to higher input temperature $T_{\rm out}$, 
 higher temperatures in the accreting flow, and stronger radiation pressure forces, because $P_{\rm out} \propto T_{\rm out}^{4} \propto {\rho}_{\rm out} {a_{\rm out}}^2$, where $P$ is the total pressure. Therefore, to establish the pressure balance at the shock front, the shock in model with the higher input density must move down to  inner location with a stronger gravitational force.

\subsection{ Overall flow and shock structure}
Fig.~3(a) shows the temperature $T$ and the Mach number
 on the equator of the steady flow in model Thick3.  In this model, the gas is optically
 thick throughout the flow except for the funnel region.
 The luminosities in models Thick2 -- Thick5 are chaotically variable with small amplitudes
 of a few factor and the chaotic variation of the luminosity is attributed to the
Kelvin-Helmholtz instability which occurs along the centrifugal barrier surface.
The averaged luminosities are
as high as $1.1 \times 10^{39}$ -- 2.7 $\times 10^{40}$ erg s$^{-1}$.
 Such powerful radiation will influence the pre-shock and post-shock structures.
 As far as 1D profiles of Fig.~3(a) are concerned,
 the shock front seems to be discontinuous similarly to the adiabatic model AD, and
 the large effects of the radiation on the shock structure seem not to be found there.
 However, Fig.~3(b)  shows an oblique shock wave with a finite thickness which
 is elongated upward from  $r \sim 10$ on the equator. The shock has a finite thickness of a few $r_{g}$ (several mesh points).
 Fig.~4 shows  2D contours of the density (a) and the temperature (b) in model Thick5.
 The overall flow consists of two regions; (1) the funnel region with an opening angle of $\sim 
 15^\circ$ between the rotational axis and the centrifugal barrier and (2) the gas accreting region between the centrifugal barrier and the equatorial plane. 
 The centrifugal barrier is roughly defined by a balance between the centrifugal force and the gravitational force.
 The accreting region is separated into further two regions by the oblique shock wave.
 The finite shock thickness is due to the radiative effect.
The figure shows the turbulent phenomana along the centrifugal  barrier and the turbulent 
gas becomes optically thick in the funnel region with increasing input density.  The turbulent motion is originated
in the Kelvin-Helmholtz instability which can occur when there exists a velocity shear in a single continuous fluid or if there is a velocity
 difference across the interface between two fluids \citep{b30}.  We confirm here the upward moving gas and the downward accreting gas across the centrifugal barrier above $z \ge 10$ in the funnel region. 
 In these models with high mass accretion rates and luminosities, the accretion disc is optically and geometrically thick
 and  it may be unfavourable for us to name the flow as ``accretion disc" because $h/r$ $\geq 1$ 
 in the inner region of the flow, if we define the disc height to be a height at which the 
  density drops to one-tenth of the central density on the equatorial plane.

\begin{figure}
 \centering
 \subfloat[2D contours of  density  (g cm$^{-3}$) in model Thick5.
  The turbulent motion appears in the funnel region along the centrifugal  barrier surface and 
  the accreting region is separated into two region by the oblique shock front elongated
   upward at $r \sim 8$]{
 \label{fig4a}
 \includegraphics[width=86mm,height=66mm]{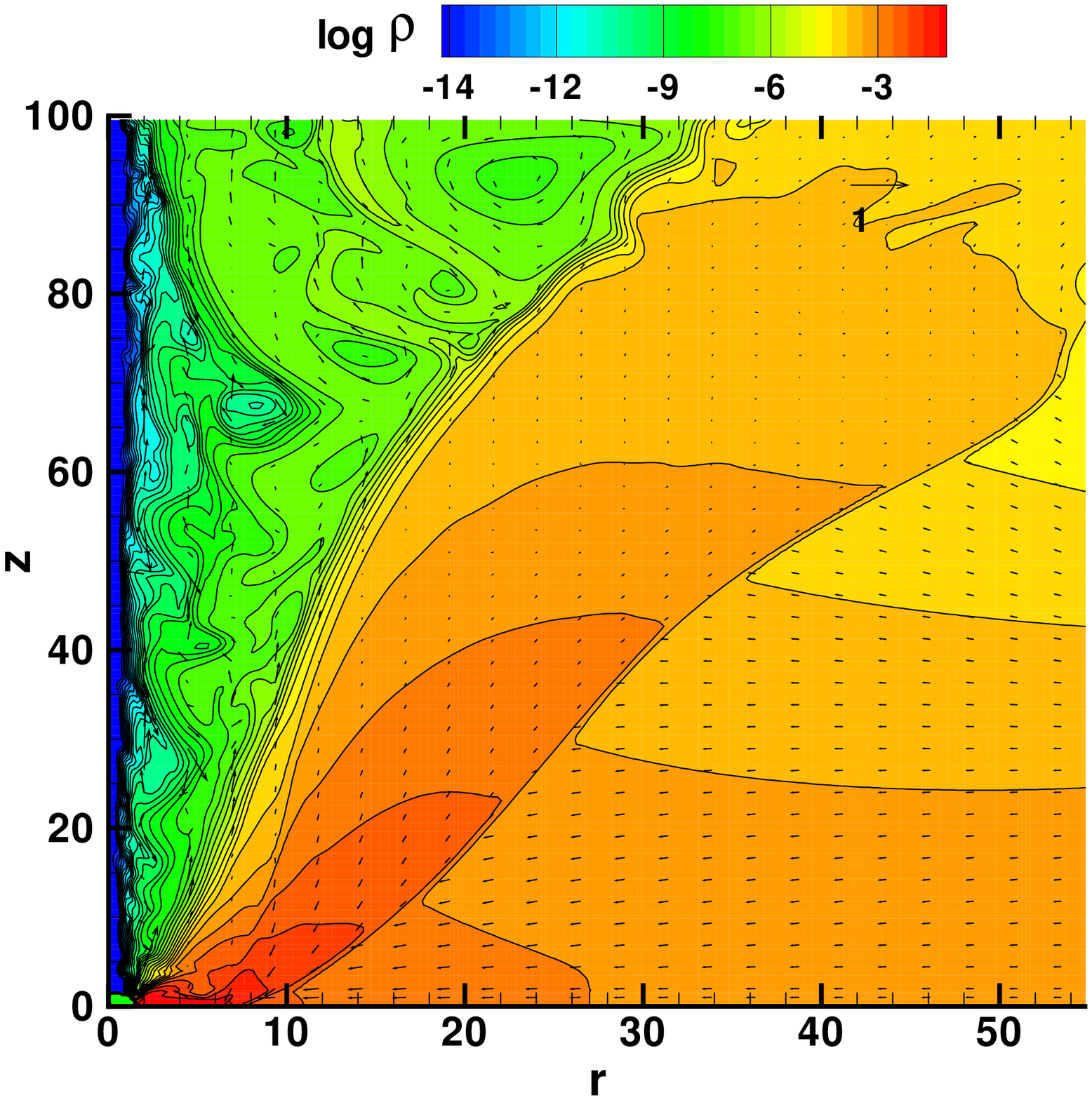} } 
 
\subfloat[Same as Fig~4(a) but for temperature in model Thick5.]{
 \label{fig4b}
 \includegraphics[width=86mm,height=66mm]{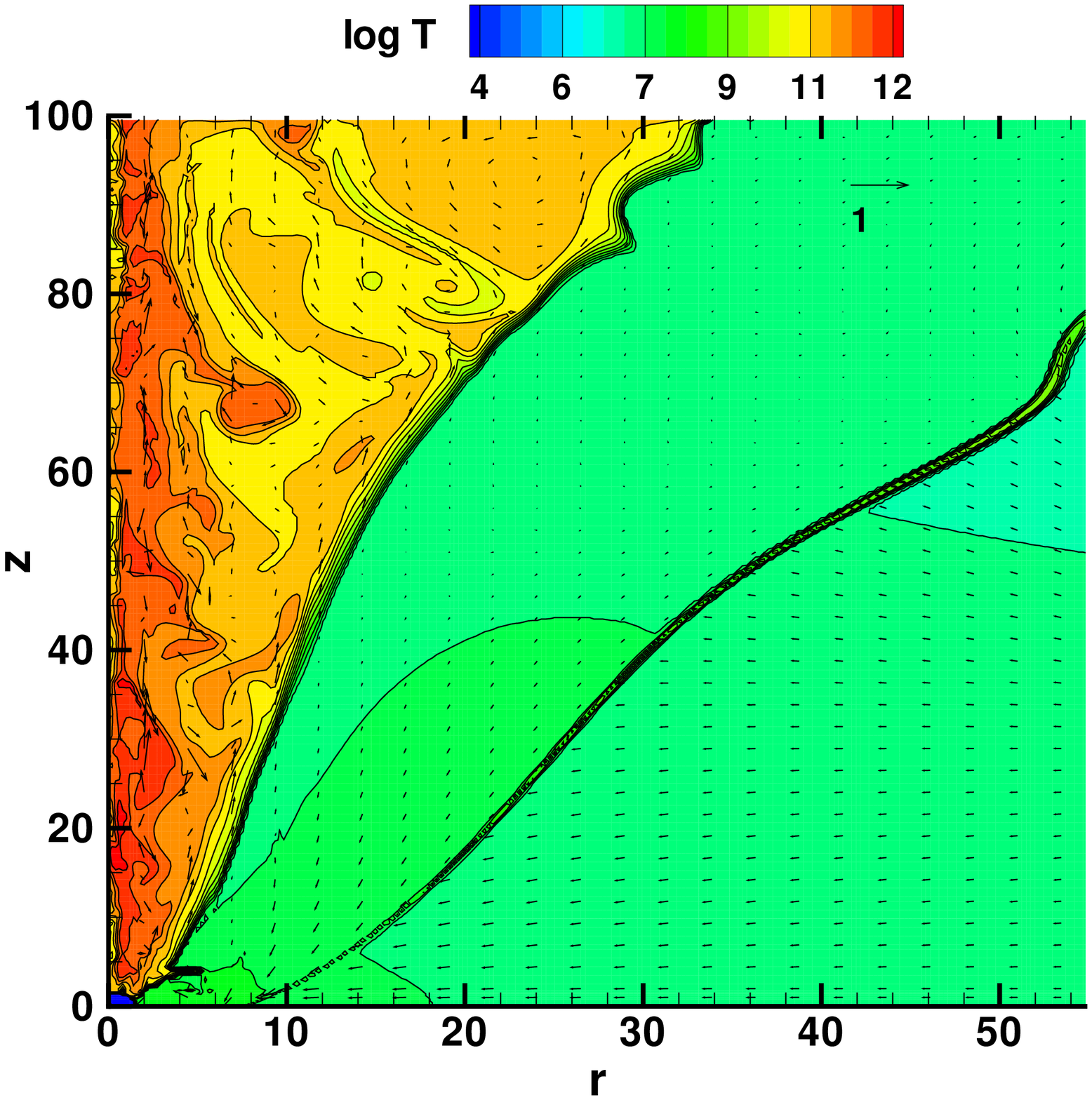} } 
 \caption{}
 \label{fig4}
 \end{figure}

\begin{figure}
 \begin{center}
     \includegraphics[width=76mm,height=66mm]{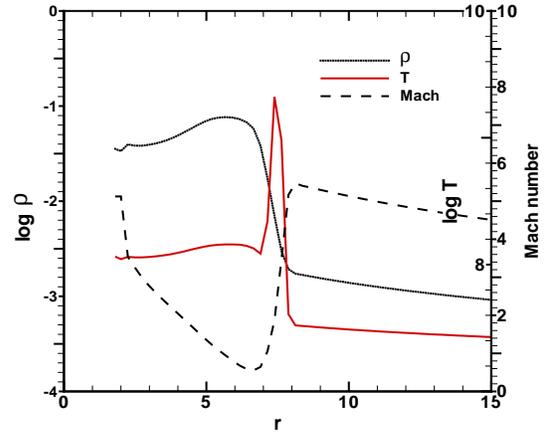}
   \caption{  Profiles of temperature $T$(K) (solid line), density $\rho$ (g cm$^{-3}$) (doted line) and  Mach number (dashed line) on the equatorial plane in model Thick5.}
  \label{fig5}
 \end{center}
 \end{figure}

  \begin{figure}
 \begin{center}
     \includegraphics[width=76mm,height=66mm]{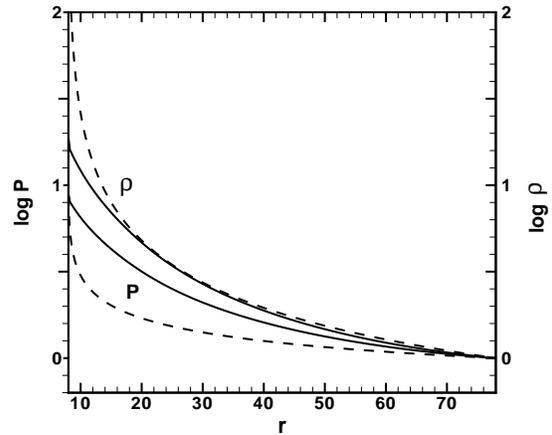}
     \label{fig6}
 \end{center}
 \caption{Comparison of total pressure and density in the precursor regions by  simulations (solid lines)
 and  analytical results (dashed lines), respectively,  where the shock front location $R_{\rm s}$=8 and 
 the upstream point in the calculation is taken as $r=78$ (see Fig.~5).}
 \end{figure}

 Fig.~5 shows the temperature $T$, the density $\rho$, and the Mach number on the equatorial plane in model Thick5, where 
  a spike-like feature of the temperature across the shock front is found,  differently from 
 the discontinuous front in the optically thin models AD and Thin1.
Under consideration of the radiation heat conduction and the supercritical shock wave, 
 \citet{b41} estimates the optical thickness $\Delta \tau_{\rm s}$ of the shock region as,

 \begin{equation}
   \Delta \tau_{\rm s}  = {4 \over {3\surd{3}}} \left[ \left({T_{-} \over T_{\rm c}}\right)^3 -1\right]  ,
 \end{equation}
 where  $T_{-}$ and $T_{\rm c}$ are the temperature before  the shock discontinuity and 
 the critical temperature in the supercritical shock, respectively.
From definition of $T_{\rm c}$ and numerical value $T_{-}$ in the simulations, we estimate
  $T_{-}/T_{\rm c} \sim $ 2.5 and then $\Delta \tau_{\rm s} \sim 10$.
 The shock thickness $\Delta R_{\rm s}$ is given by $\Delta R_{\rm s}$ =  $l_{\rm r} \:
 \Delta \tau_{\rm s}$, where $l_{\rm r} = 1/{\kappa \rho}$ is the phton mean free path and 
 we use  the Kramers opacity for $\kappa$.
 Then we have $\Delta R_{\rm s} \sim 7 \times 10^6$ cm $\sim$ a few $r_{\rm g}$.
The shock thickness agrees well with $\sim$ a few $r_{\rm g}$ obtained in the simulations of 
models Thick 3 to 5. 
 In model Thick5, the gas is
 fully optically thick and radiation-pressure dominant  in the pre-shock and post-shock regions. But, in the  peak temperature region behind the shock front, the gas is moderately optically thick and the ratio of gas pressure to total pressure is  $\sim$ 0.1.

 \citet{b16} examined  non-relativistic radiative shocks in the disc accretion and solved analytically the structure of  the radiative precursor region for both the gas
 and radiation-pressure dominated cases.
 Since our model Thin1 with gas-pressure dominant and optically thin matter shows almost the same result 
 as the adiabatic case, the radiative effects on the shock are negligible.
 We compare the analytical structure of the precursor region in the pressure dominant flow
 with the results by the simulations for model Thick5.
 The structure equation for the radiative precursor in the radiation-pressure dominated case is given as \citep{b16},

  \begin{equation}
    { 1 \over {\tau_{1} \beta_{1}} }  \frac {\tilde{P}^2} {\tilde{\Sigma}^3}  \frac {d \tilde{P} } {d x}  =
       {1\over 2} {\gamma \over {\alpha_{1}}}
      M_{1}^2 ({1\over {\tilde{\Sigma}^2}}-1) + 4 ({\tilde{P}^2\over {\tilde{\Sigma}^2}} - 1),  
  \end{equation}

  where

  \begin{equation}
   \tilde{\Sigma} = \frac {\tilde{P}^2 + ({\gamma\over {\alpha_{1}}} ){M_{1}}^2} {1 +  ({\gamma \over {\alpha_{1}}} ) {M_{1}}^2}.
  \end{equation}
  Here $\tau_{1}$, $\beta_{1}$, $\alpha_{1}$ and $M_{1}$ are the parameters determined at the upstream region far from
 the shock, $\tilde{P}$ and $\tilde{\Sigma}$ are the total pressure and the surface density normalized in their values at the upstream region.
 With $\tau_{1}=4.3\times 10^{-3}$, $\beta_{1}=8.6\times 10^{-2}$, $\alpha_{1}= 2\times 10^2$, $M_{1}= 2.77$, and other
  variables at the upstream region in model Thick5, we get the analytical solutions of  $\tilde{P}$, $\tilde{\rho}$, $\tilde{\Sigma}$,
 etc.,  solving  the equation (17) by Runge-Kutta method. 
 Fig.~6 shows the solutions of $\tilde{P}$ and $\tilde{\rho}$ in the precursor region of model Thick5 (solid lines) and the analytical solutions (dashed lines), respectively, where shock front location $R_{\rm s} = 8$ 
and the upstream point in the calculation is taken as $r = 78$.
 The analytical density is higher more than two orders of magnitude than 
 the density in the simulation only near the shock front and that the
  pressure is underestimated   compared with that in the simulations.
 Since $P \propto E_{\rm 0} \propto T^4$ in the optically thick and radiation-pressure dominant region, the simulation shows considerable heat up of the upstream region.
 The theoretical solutions are based on the assumption of the vertically hydrostatic equilibrium
 which requires the geometrically thin disc as $h/r \ll 1$.  However, when the mass accretion is sufficiently large and the luminosity exceeds the Eddington luminosity, the accretion disc becomes geometrically thick.
Our simulations in models Thick4 and 5 show $h/r \approx 1$ in the inner region of the flow. 
 Then, the geometrically thick and hot disc leads to a lower density.
However, there is no big difference between the theoretical and the simulations results for the present model. This may be due to the small Mach number $\sim 2$ used here.

\subsection{Astrophysical relevance}
In models Thick4 and 5, the mass outflow rates are a few percent of the input accretion rate but the absolute rates are as high as $\sim 10^{-5}$ -- $10^{-4}$ $ M_{\odot}$ yr$^{-1}$. 
 About 80 percent of the mass outflow is lost from the outer radial
 boundary and the mass  outflow rate from the funnel region is small as one thousandth of
 the total  outflow rate but the absolute flow rate is rather high as $\sim 10^{-6} 
 M_{\odot}$ yr$^{-1}$ compared with usual mass loss rate of normal stars.
 The high mass outflow rate may be observed over the wide wind region because the 
  outflow from the radial outer boundary will develop as the wind.
 Besides,  the radial velocities on the vertical outer boundary within the funnel region  are 
  small as $\sim$ 0.08 maximumly, which is smaller than the escape velocity $\sim 0.1$, and somewhere negative because of the turbulent motion.
 Then, the high velocity jets are not found here.

As is found in Table 2, the luminosity  $L_{\rm f}$ through the funnel region covers  $\sim$
80 percent of the total luminosity of 1.2 -- 2.7 $\times 10^{40}$ erg s$^{-1}$ in models Thick4 and 5. 
 While the luminosity $L_{\rm r_{\rm out}}$ from the outer radial boundary 
 is very small as  $\sim 10^{-4}$ of the total luminosity.
 This means that the face-on luminosity is very high as $\sim 10^{40}$ erg s$^{-1}$ but
 the edge-on luminosity is small as $\sim 10^{36}$ erg s$^{-1}$.  
  Fig.~7 shows the  luminosity distribution  $\Delta L(r)$ per area over 10 $r_{\rm g}$ on the 
  vertical outer boundary surface in models Thin1, Thick4, and Thick5.
  It shows a strongly anisotropic distribution of radiation around the rotational axis
 in models Thick 4 and 5 but not in optically thin model Thin1.
Why the luminosity  from the funnel region is so high ? In the usual funnel region 
under the accretion disc flow, gas is very rarefied, optically thin and hot. 
 Actually, in models AD, Thin1, and Thick2, the gas in the funnel region is optically thin and hot but partly optically thick in model Thick3.
 On the other hand, in models Thick4 and 5, the input densities are much higher.
 The Kelvin-Helmholtz instability occurring along the centrifugal barrier transports upward turbulent  outflow gas from the base of the inner accretion disc. As a result, the gas 
in the upper-funnel region  becomes dense and optically thick with the increasing input density. 
The existence of the radiative shock promotes such a process in the inner region because the shock is formed near to the event horizon. Thus, the optically thick and hot gas in the funnel region contributes largely to the total luminosity with an increasing mass accretion rate.
 The optically thick portion in the funnel region widens with the increasing input density and 
 the ratio  $L_{\rm f}/L$  of the funnel luminosity to the total luminosity in Table 2 shows  such result.

\begin{figure}
 \begin{center}
  \begin{tabular}{cc}
  \begin{minipage} {0.3\linewidth}
   \begin{center}
     \includegraphics[width=25mm,height=36mm]{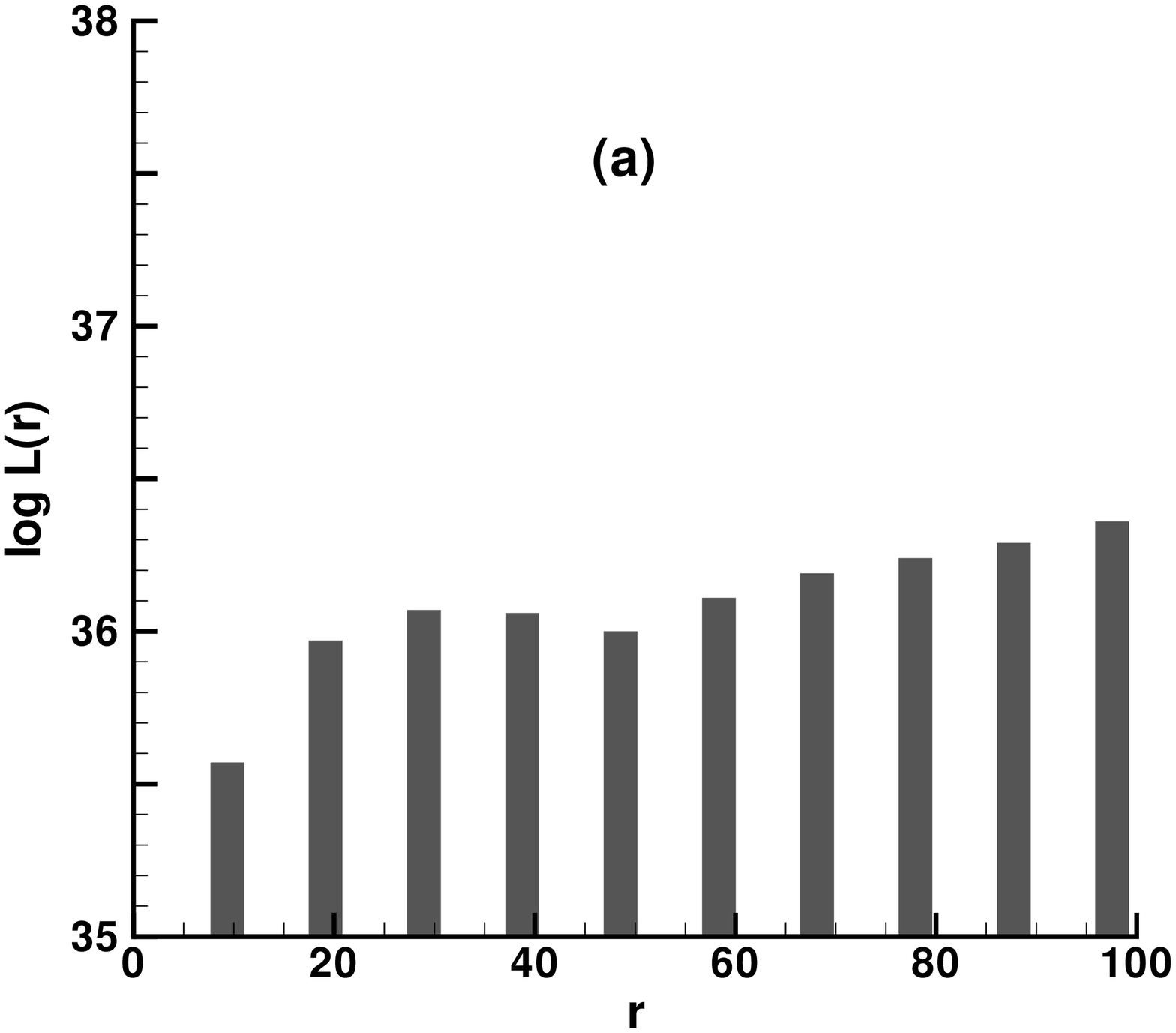}
     \label{fig7a}
 \end{center}
 \end{minipage}
%\hskip 0.5cm
\begin{minipage}{0.3\linewidth}
  \begin{center}
   \includegraphics[width=25mm,height=36mm]{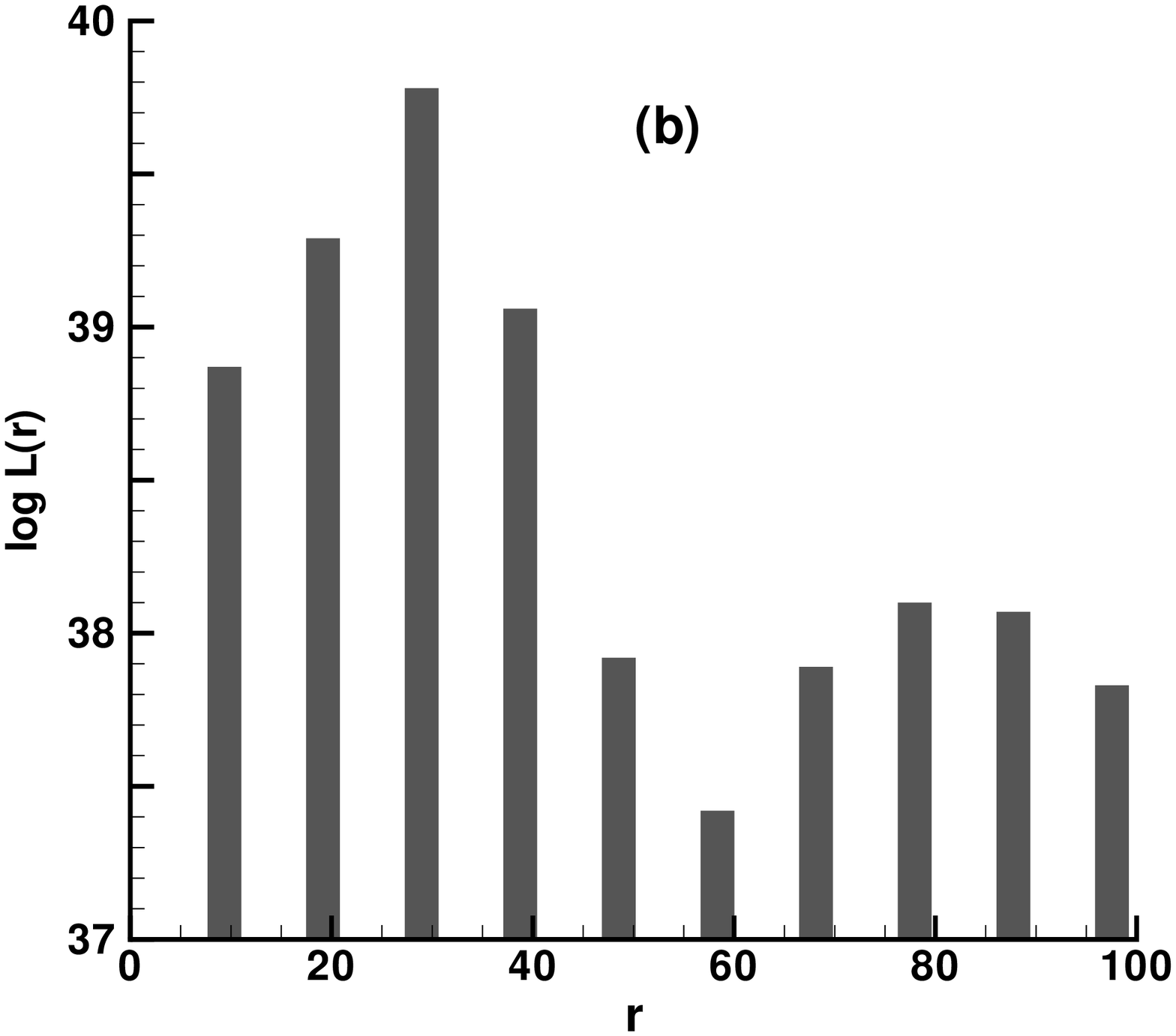}
   \label{fig7b}
  \end{center}
  \end{minipage}
%\hskip 0.5cm
\begin{minipage}{0.3\linewidth}
  \begin{center}
   \includegraphics[width=25mm,height=36mm]{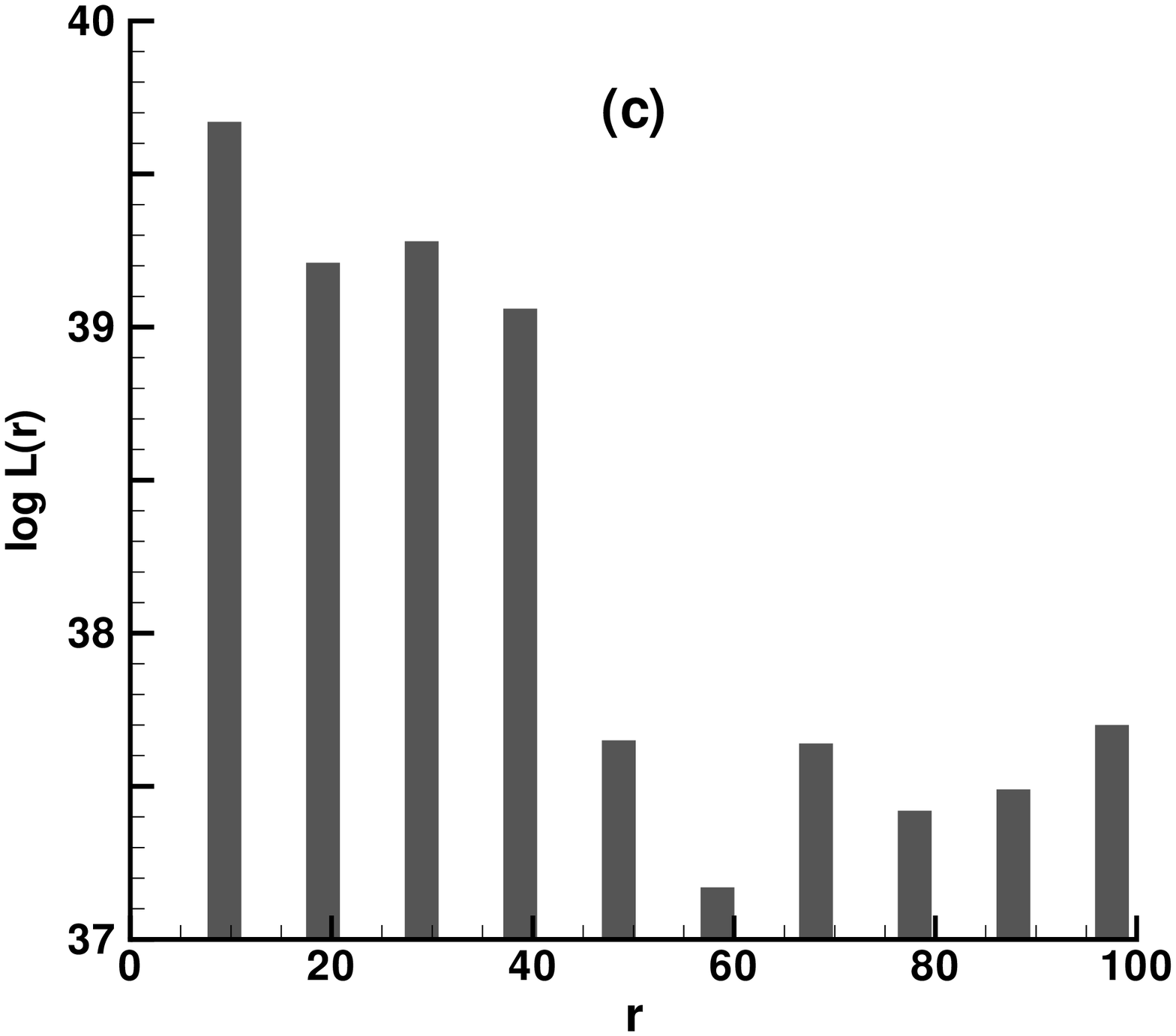}
   \label{fig7c}
  \end{center}
  \end{minipage}
  \end{tabular}
  \caption{Radial distribution $\Delta L(r) $ (erg s$^{-1}$) of  luminosity per area over
 10$r_{\rm g}$ on the
  vertical outer boundary surface in models Thin1 (a), Thick4 (b) and Thick5 (c).
 Most of the radiation in models Thick4 and 5 is emitted from the optically thick funnel
 region within $r \le 35$ but optically thin  model Thin1 never show strongly anisotropic 
 distribution of the radiation. } 
 \end{center}
 \end{figure}

The very high luminosity from the narrow funnel region, the low edge-on luminosity, and
 the  high mass outflow rate from the wind region are characteristic features of the optically thick models considered here. In these models, the gas is fully optically thick in the accretion flow and  optically thick even in the funnel region except just near the rotational axis.
 For simplicity, assuming that the optically thick gas on the vertical outer boundary emits as a black body but only optically thin funnel region radiates the free-free emission,
 we get a rough spectrum $L_{\nu}$ given by

\begin{equation}
  L_{\nu} = \int \pi B_{\nu}(T) d \textbf{S} + \int \epsilon_{\rm ff}(\nu) {\rm d} V
       \;\; {\rm erg}\; {\rm s}^{-1}\; {\rm HZ}^{-1},
 \end{equation}
where $B_{\nu}(T)$ is the Planck function, $\epsilon_{\rm ff}$ is the monochromatic free-free emission rate, and the area  and  volume integrals are done over  the optically thick surface on the vertical outer boundary and the funnel region, respectively.
Fig.~8 shows the spectra $L_{\nu}$ for models Thick4 and 5. 
 The black body spectra with  a single temperature of  $T =5\times 10^{6}$ (dotted line) and  $10^{7}$ (dashed line) K in model Thick4 and 
 $T =10^{7}$ (dotted line)  and  $3\times 10^{7}$ (dashed line) in model Thick5 are also shown, respectively.
The spectra in models Thick4 and 5 behave approximately as the black body radiation with a single
 temperature of $T=5\times 10^6$ -- 
 $3\times 10^7$ K.

\begin{figure}
 \centering
 \subfloat[Spectra $L_{\nu}$ erg s$^{-1}$ Hz$^{-1}$  for model Thick4 (solid line), where
  dotted and dashed lines denote black body radiations with a single temperature of 
 $T = 5\times 10^6$  and $10^7$ K,  respectively.]{
 \label{fig8a}
 \includegraphics[width=76mm,height=56mm]{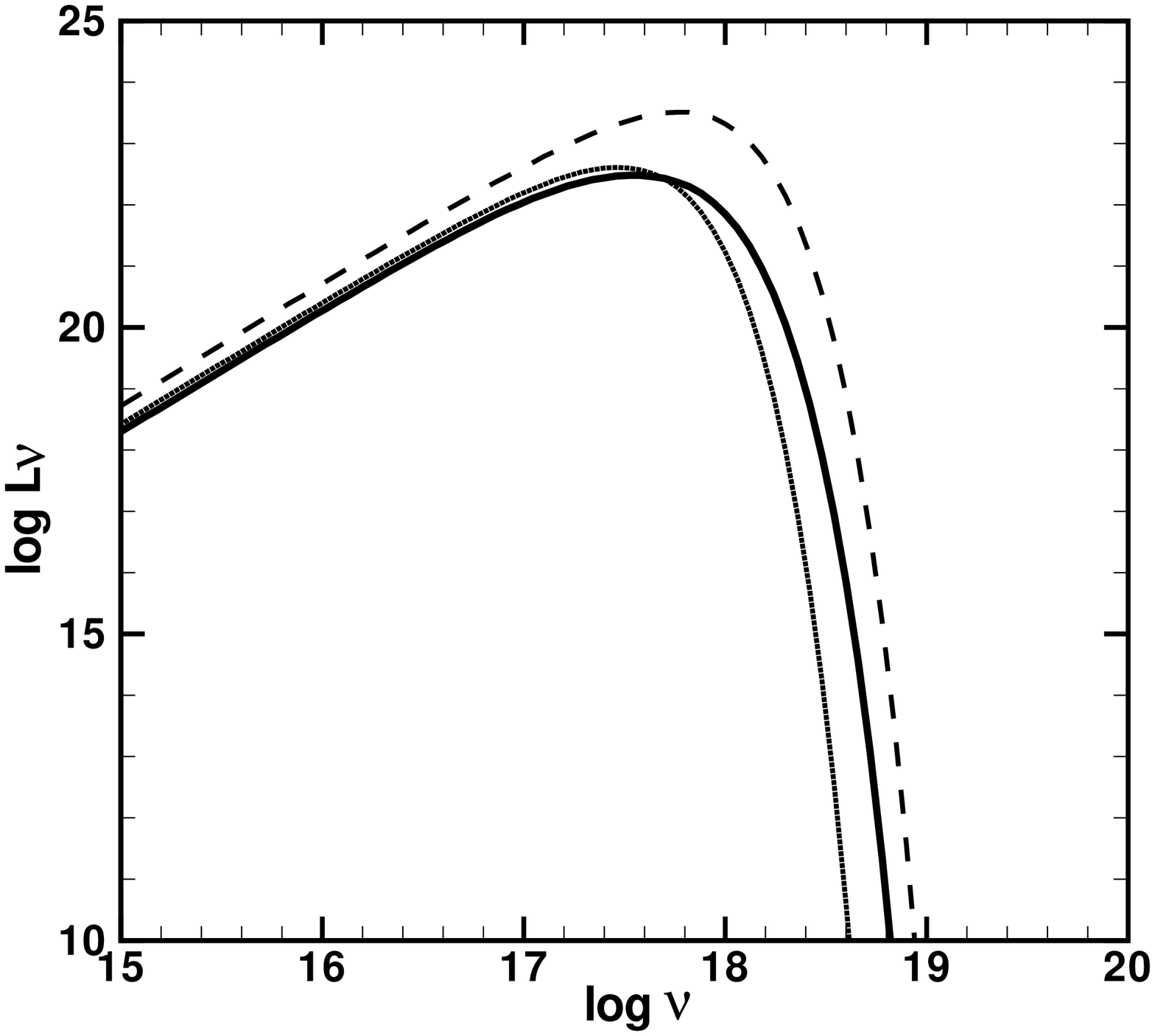} } 
 
\subfloat[Same as Fig.~8(a) but in model Thick5 (solid line) and  black body radiation with
 a single temperature of  $T = 5\times 10^6$ (dotted line)  and $3\times 10^7$ K (dashed line). 
 The free-free emission component from the funnel region is almost negligible here]{
 \label{fig8b}
 \includegraphics[width=76mm,height=56mm]{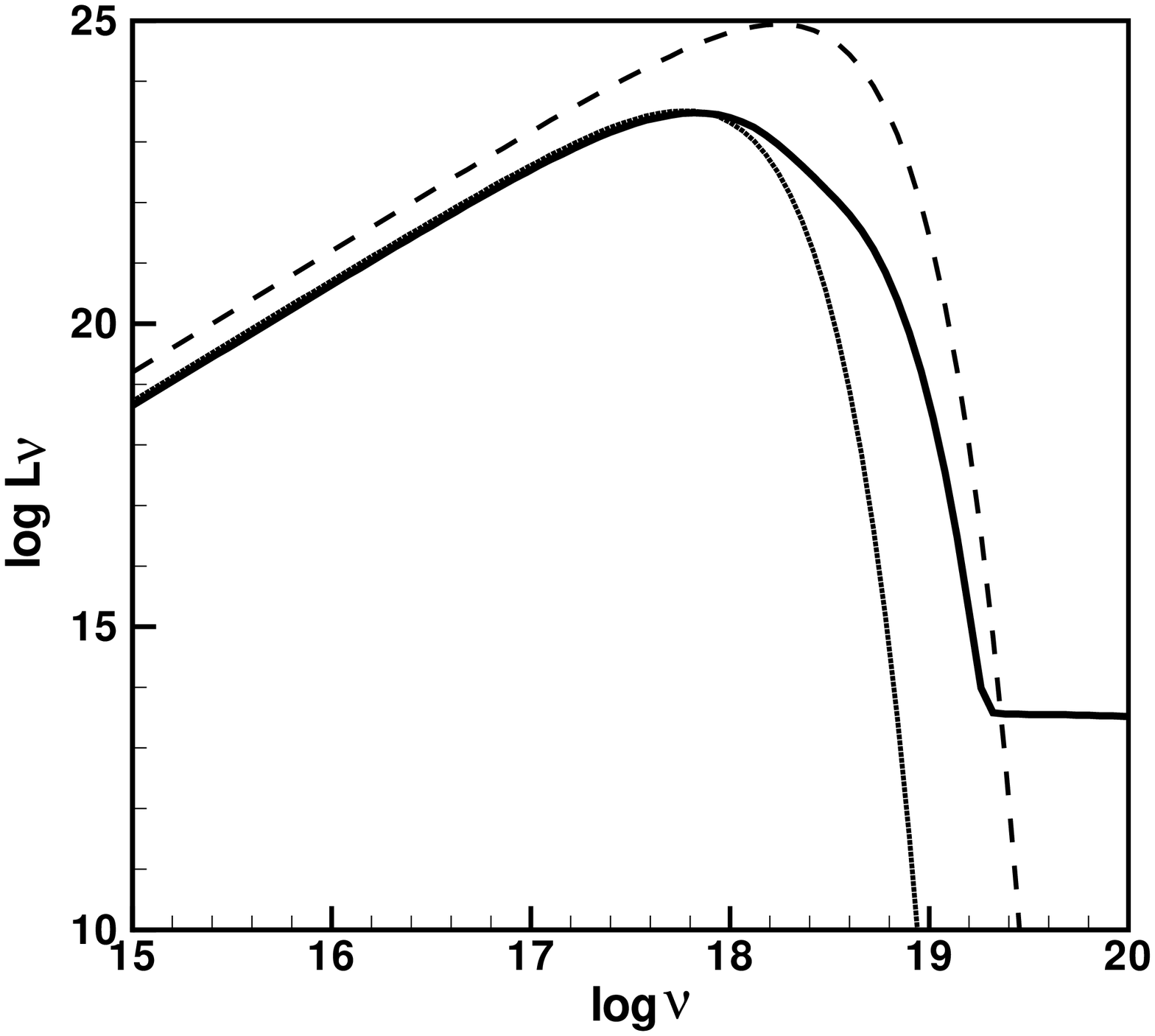} } 
 \caption{}
 \label{fig8}
 \end{figure}

\section{Summary and Discussion}
We studied  radiative shocks in low angular momentum flow around super-Eddington accreting
 black holes.
Adopting a typical set of  flow parameters of the specific angular momentum $\lambda$, the radial velocity $v_{\rm out}$, the sound velocity $a_{\rm out}$, and the density $\rho_{\rm out}$  
 responsible for the standing shock,
 we obtained  the results which are summarized below.

 (1) In the optically thick and radiation-pressure dominant models, the flows
 with input accretion rate of $1.2 \times 10^2$ to $1.2 \times 10^5$ Eddington critical accretion rates result in 
 $\sim$ 1 -- 18  $L_{\rm E}$ ($ 10^{39} $ -- $ 3 \times 10^{40}$
 erg $s^{-1}$) and the luminosities tend to be saturated towards $\sim 3 \times 10^{40}$ erg s$^{-1}$
 with increasing accretion rates. The shock locations $R_{\rm s}$ on the equator 
  decrease as $\sim$15 to 8 with the increasing accretion rate, compared with the large $R_{\rm s} \sim 50$ in the adiabatic model with the same parameters. 
 This is because the increasing mass accretion rate leads to higher temperatures and
 stronger radiation pressure forces in the accreting gas and that, to establish the pressure balance at the shock front, the shock with the higher accretion rate must move down to the inward location with stronger gravitational force. 

 (2)  The shock wave is formed obliquely from the equator, and the oblique shock has a finite
 thickness of the shock front due to the radiative effect. The finite shock thickness
 $\Delta R_{\rm s}$ of a few $r_{\rm g}$ corresponds well to  theoretical value
  by Zel'dovich and Raizer (1966).  
 The shock structure in the precursor region agrees with the theoretical one by 
 Fukue (2019c), except just near the shock front

(3)  In optically thick models Thick4 and 5, the mass outflow rates are high as $\sim
 10^{-5}$ and $10^{-4}  M_{\odot}$ yr$^{-1}$ and are mostly originated in the wide wind region. 
  Most of the radiation  $\sim 10^{40}$ erg $s^{-1}$ is emitted from the narrow funnel region 
 along the rotational axis, that is,  the radiation shows a strongly anisotropic distribution around 
 the rotational axis. This is because the gas in the funnel region becomes optically thick
 and hot with increasing input accretion rate and emit radiation like the black body with temperatures 
 of  $5\times10^6$ -- $3\times 10^7$ K.  While the edge-on luminosity is far small as $\sim 10^{36}$ erg $s^{-1}$. 

 The present optically thick and radiation-pressure dominant shock models show some of the characteristic features of Ultraluminous X-ray sources (ULXs). ULXs are X-ray sources with X-ray luminosity above the Eddington limit for stellar-mass or below the  Eddington limit for intermediate-mass black
holes but not supermassive black holes \citep{b39, b40, b2},
 although the recent discovery of coherent pulsations in ULXs support 
 neutron stars as host \citep{b3}. 
 A super-Eddington accretion model onto a stellar-mass
 black hole whose luminosity far exceeds the Eddington luminosity was proposed by  
 \citet{b37}. The key features of the super-Eddington accretion are strong optically thick wind and collimated radiation. 

 Such super-Eddington models were well reproduced by 2D radiation hydrodynamic
and 2D radiation magneto-hydrodynamic simulations, which take account of viscous and magnetic dissipation, respectively \citep{b29, b28}.
 They found that the matter can fall onto the black hole with an accretion rate beyond the Eddington limit and the apparent luminosity can exceed the Eddington luminosity because of the photon trapping within the disc. As a result, the supercritical accretion flows show mildly anisotropic and collimated emission and a strong outflow from the disc due to radiation pressure force. The basic difference between their and our work is that we consider the inviscid, low angular momentum flow while they focus on the viscous, high angular momentum flow.

The superaccretor SS 433 belongs to a typical class of ULXs. The apparent X-ray luminosity is about $\sim 10^{36}$ erg s$^{-1}$ but the intrinsic luminosity is considered to be probably
  $\geq 10^{39}$ erg $s^{-1}$ because we observe the source edge-on (Fabrika 2004, Fabrika, Vinokurov and Atapin 2018).
In our models Thick4 and 5, the luminosities are very high as $\sim 10^{40}$ erg s$^{-1}$ and 
the radiation shows a strongly anisotropic distribution around the rotational axis.
 In addition, the edge-on luminosity through the outer radial boundary is very low as $\sim 10^{36}$ erg s$^{-1}$. 
The large mass outflow rates $\dot M_{\rm out} \sim 10^{-5}$ -- $10^{-4}  M_{\odot}$
 yr$^{-1}$ are originated in the wide wind region. However, the relativistically  high velocity jets
 along the rotational axis are not found in the funnel region, as is detected in SS 433. This may be due to that the outer boundary used here is small as $R_{\rm out}$ = 100
 and the outflow gas in the funnel region is not still sufficiently accelerated by powerful radiation 
 pressure force.
 In SS 433 which consists of a close binary system, 
 the mass transfer rate from the second companion  and the wind mass outflow rate from the primary 
 are observationally estimated to be  $\sim 10^{-4}$ -- $10^{-3} M_{\odot}$ yr$^{-1}$ 
and  $\sim 10^{-5}$ -- $10^{-4}  M_{\odot}$ yr$^{-1}$, respectively \citep{b11}.
 Their values agree  well to the input mass
 accretion rates and the mass outflow rates in models Thick4 and 5. 
 In the close binary system with high luminosity, it is conceivable that the companion
 transports not only high mass outflow but also high angular momentum to the primary through critical
 Roche lobe, differently from small angular momentum by stellar wind in detached binary system.
Then, one may wonder if the inviscid flow used in this paper is valid under such close binary system with the super-Eddington luminosity or not.
This is justified as the viscous flow can achieve specific angular momentum of Keplerian value at the outer boundary while behaving as a sub-Keplerian, low angular momentum 
 flow with shock in the inner region \citep{b21}. Actually they find the global accretion solution that  the viscous Keplerian flow with the viscosity parameter $\alpha$ ($\le$ 0.15) at the outer boundary beyond distance of  $5 \times 10^3 r_{\rm g}$ connects to the low angular momentum sub-Keplerian flow below the inner region of a few hundreds $r_{\rm g}$.

 Finally, the radiative shock models give a concrete shape to the superaccretor SS 433  and show a possible model for ULXs.
Further studies of the radiative shock model with the super-Eddington luminosity
 should be developed to apply to ULXs with stellar-mass black holes.

\section*{Acknowledgments}
CBS is supported by the National Natural Science Foundation of China under grant
no. 12073021.

\section*{Data Availability}
The data underlying this article will be shared on reasonable request to the corresponding author.

\label{lastpage}

\end{document}